\documentstyle[12pt,epsf]{article}

\newcommand{\eq}[1]{\begin{equation}\label{#1}}
\newcommand{\eqs}{\begin{equation}}
\newcommand{\en}{\end{equation}}
\newcommand{\bea}{\begin{eqnarray}}
\newcommand{\ena}{\end{eqnarray}}
\newcommand{\r}{{\bf r}}
\newcommand{\Z}{{\bf Z}}
\newcommand{\X}{{\bf X}}
\newcommand{\h}{{\bf h}}
\newcommand{\U}{{\bf u}}
\begin{document}
\title{Multiscaling in passive scalar advection as stochastic shape dynamics}
\author{Omri Gat(1) and Reuven Zeitak(1)(2)\\
(1)Dept. Chem. Physics, The Weizmann Institute of Science,\\
 Rehovot 76100, Israel\\
(2) Laboratoire de Physique Statistique, ENS,\\
 24 rue Lhomond, 75231 Paris Cedex 05, France\\
}
\maketitle
\abstract
{ The Kraichnan rapid advection model is recast as the stochastic
dynamics of tracer trajectories. This framework replaces the random
fields with a small set of stochastic ordinary differential equations.
Multiscaling of correlation functions arises naturally as a
consequence of the geometry described by the evolution of $N$
trajectories. Scaling exponents and scaling structures are interpreted
as excited states of the evolution operator. The trajectories become
nearly deterministic in high dimensions allowing for perturbation
theory in this limit. We calculate perturbatively the anomalous
exponent of the third and fourth order correlation functions. The
fourth order result agrees with previous calculations.

}

\subsection{Introduction}

Although most researchers in the field of turbulence agree that
the structure functions of the velocity field exhibit multiscaling,
there is no well understood mechanism for this phenomenon which would
allow, even in principle, a systematic calculation of the values of
the multiscaling exponents \cite{legacy}.
The Kraichnan rapid advection model \cite{kr68} is a simplified model for
turbulent advection of a passive scalar in which a mechanism for
multiscaling has been identified. Although the relevance of this mechanism to
multiscaling in fully developed turbulence in still unclear, the study
of multiscaling in Kraichnan's model
presents a new approach and therefore has attracted a lot of interest
recently \cite{kr94,gk,grisha,3pt}.

One of the main advances in studying this model was the identification
of anomalous scaling contributions to correlation functions that are
zero modes of certain linear partial differential operators
\cite{grisha}. Although this
identification has allowed for the development of perturbation
theories for the computation of the anomalous exponents, the meaning
of these structures is, in our view, still somewhat obscure. The
purpose of this paper is to reinterpret the anomalous exponents as
eigenvalues of an evolution operator for the relative shape of
trajectories of particles advected by the flow. The zero-modes arise
naturally as the
adjoints of the set of eigenfunctions of this evolution.

Our method is to consider the evolution of $N$ particles with the
flow. These $N$ particles define a configuration that has an overall
scale and a normalized shape. Under the flow dynamics, the overall
scale
tends to increase while the shape will asymptotically be described by
a stationary measure. We consider the rate at which
a distribution of initial shapes tends towards its stationary
distribution. It is important to consider this rate in terms  of the
overall scale. Doing this
we find the decay rate towards the stationary state to be a
combination of power laws in this scale. We identify
these power laws with the anomalous exponents of the passive scalar.
Thus we view the anomalous scaling as a dynamic
property of the evolution of trajectories \cite{slow}.

This point of view treats the anomalous scaling as a
geometric property of the trajectories of fluid particles in the flow,
without explicit reference to the forcing. In fact, one does not need
to define a passive scalar field since all the properties we are
interested in are included in the shape dynamics. Note that this explains why
the two point correlation functions do not have anomalous scaling --- the
geometry of a two point configuration is trivial, since it is completely
described by the distance between the points.

We will show that this picture formally contains all the ingredients of
anomalous scaling in the dynamics of passive scalars. Furthermore, it
is useful as a calculational scheme. We will demonstrate this by
presenting an alternative derivation for perturbation theory in large
dimensions. We believe that the physics of passive scalar advection in
large dimensions becomes more transparent in this framework. We
calculate the previously unknown perturbative correction to the
anomalous exponent of the third
order correlation function, and recover the known result for the
fourth order case \cite{grisha}.

In section 2, we present the path integral formalism of passive scalar
advection, and show how correlation functions can be written as
averages over Lagrangian trajectories. In section 3 we exploit this
picture to identify a mechanism for generation of anomalous scaling,
and show how the anomalous exponents are related to eigenvalues of
an evolution operator. In section 4 we apply this picture to calculate
perturbatively the first correction to the anomalous
exponents in large dimensions. Section 5 is devoted to summary and
conclusions.

\subsection{Lagrangian path formulation of passive scalar advection}

The dynamics of a passive scalar $\theta$ advected by an incompressible
velocity field $\U$ are described by
\eq{passive}
\partial_{t} \theta(\r,t)=\kappa \nabla^{2} \theta(\r,t)-
\U\cdot \nabla \theta(\r,t)+f(\r,t)
\en
where $\kappa$ is the molecular diffusivity, and $f$ models the
injection and extraction of scalar by external sources. The properties
of $\U$ and $f$ are given, presumably as the result
of some physical process.

In the rapid advection model \cite{kr68} the velocity field is taken
to be a random
Gaussian field, delta correlated in time and self similar correlations
in space:
\eq{vel} \left< (\U(\r,t)-\U(\r',t))\otimes(\U(\r,t')-\U(\r',t')) \right> =
\h(\r-\r') \delta(t-t'), \en
where the ``eddy-diffusivity'' tensor $\h$ is defined by
\eq{eddy}
\h(\r)=\left({r\over \ell}\right)^\xi
({\bf1}-{\xi\over d-1+\xi}{\r\otimes\r\over r^2})\ ,
\en
where the coefficients are chosen such that $\nabla\cdot\h=0$.
The forcing $f$ is also taken white in time, Gaussian, and is
characterized by a single large scale $L$,
\eq{forc} \left<f(\r,t)f(\r',t')\right>=\Xi(\r-\r') \delta(t-t'), \en
where the function $\Xi(r)$ is nearly constant for $r\ll L$ and
decays rapidly for $r>L$.

The rapid advection model is theoretically attractive due to the fact
that the equations for the correlation functions are linear partial
differential equations as opposed to the hierarchy of nonlinear
integro-differential equations for fully developed turbulence.  Some
theoretical progress has been made in the case of the rapid advection
model and it is predicted that the $2n$th order correlation function
behaves as
\eq{pheno1}
F_{2n}(\r_1,\ldots,\,\r_{2n})= L^{n(2-\xi)}(c_0+\cdots+
c_k(r/L)^{\zeta_{2n}} \hat F_{2n}(\hat \r) +\cdots)\ ,
\en
where $r\ll L$ is a typical distance between points, and $\hat\r$
denotes a set of dimensionless variables describing the configuration
of the $2n$ points. According to the existing theory, the terms in the
expansion (\ref{pheno1}) are of two types: Either they are
inhomogeneous terms, whose scaling is determined by dimensional
analysis, or they are the homogeneous solutions of moment equations,
whose  scaling
exponents are not constrained by dimensional arguments. The exponents
and scaling functions are expected
to be universal, but not the $c$ coefficients, which depend on the
details of forcing.

The exponent $\zeta_{2n}$ which dominates when considering structure
functions
\eq{pheno2}
S_{2n}(\r_1,\r_{2})=<(\theta(\r_{1},t)-\theta(\r_{2},t))^{2n}>
\en
is the scaling of the leading term in (\ref{pheno1}) which depends
non-trivially on all of the $2n$ variables. Anomalous scaling occurs
when this term is due to homogeneous solutions.

As the path integral formalism for equation (\ref{passive}) is well
documented \cite{drummond}, we will not detail it here. The main
conclusion is that one may write its solution formally

\eq{basic3}
\theta(\r,t)=\int_{-\infty}^{t}dt' <f(\r'(t'),t')>_{\eta}
\en
with the trajectory $\r'$ obeying
\bea
\r'(t)&=&\r\\
\partial_{t'} \r'(t')&=&\U(\r'(t'),t')+
\sqrt{2 \kappa}{\eta}(t')\label{traject1}
\ena
and $\eta$ is a vector of  zero-mean independent
Gaussian white random variables.

The physical meaning of this equation in simply understood if we consider
the $\kappa=0$ case: the trajectories $y$ are simply the characteristics
of the partial differential equation (PDE)~(\ref{passive}), namely
the Lagrangian trajectories of the flow
field $\U$. If we want to know how much scalar is at point
$(\r,t)$ we need to see what the value of $\theta$ was where this
point came from, and add the accumulating forcing along the path
traced backward in time~\cite{dz}. When $\kappa$ is positive the trajectories
are randomly perturbed
Lagrangian trajectories and the scalar $\theta$ is obtained as
an average over $\eta$. This procedure generates the diffusion term in
equation (\ref{passive}).

Using the path integral formalism, we want to  compute correlation
function of the type
\eq{corr}
F_n(\r_1,\ldots,\,\r_n)=\left<\theta(\r_1)\cdots\theta(\r_n)
\right>_{\U,f}\ ,
\en
where $\left<A\right>_{\U,f}$ denotes an average over realizations of velocity
and forcing.
Substituting  each factor of $\theta(\r_i)$  by its representation
(\ref{basic3}) we can perform the $f$ averages explicitly, yielding
\eq{f2n}\begin{array}{r}
F_{2n}(\r_1,\ldots,\,\r_{2n},t)=\langle\int_{-\infty}^{t}dt_1\cdots dt_n
\Big[\Xi(\r'_1(t_1)-\r'_2(t_1))\cdots\qquad\\
\times\Xi(\r'_{2n-1}(t_n)-\r'_{2n}(t_n))
+\hbox{permutations}\Big]\rangle_{\U,\eta}\ ,
\end{array}\en
while the odd moments vanish. Each of the trajectories $\r'_i$ obeys an
equation of the form (\ref{traject1}), where $\U$ as well as $\eta$ is
a stochastic variable, whose correlation function is given
by~(\ref{eddy}).  The procedure just described is a Langevin dynamics of $N$
particles, with correlated noise. Incompressibility of $\U$ implies that
eq.~(\ref{traject1}) should be interpreted in the Ito sense.
The corresponding Fokker-Planck
equation with respect to {\it both} $\eta$  and $\U$ is precisely the
well-known
homogeneous Kraichnan PDE for the $2n$th moment \cite{kyc}.

The representation of $F_{2n}$ given in eq~(\ref{f2n}) can be turned
into a  recursive expression by ordering the $t_1,\cdots,t_n$
integrations such that $t_1>t_2>\cdots>t_n$. Since  eq~(\ref{f2n})
contains all permutations in $\r'_{k}$, this ordering can be done
without changing the results. The times $t_2,\cdots,t_n$
can be integrated out  to give
\eq{f2n_onetime}\begin{array}{l}
F_{2n}(\r_1,\ldots,\,\r_{2n},t)=\\
\langle\int_{-\infty}^{t}dt_1
\Big[\Xi(\r'_1(t_1)-\r'_2(t_1))
F_{2n-2}(\r'_3(t_1),\cdots,\r'_{2n}(t_1))
+\hbox{permutations}\Big]\rangle_{\U,\eta}\ .
\end{array}\en
This form for $F_{2n}$ is useful because it only contains a single time
integration, so that one needs to follow a joint trajectory of $2n$
particles at the same time and integrate along this trajectory the forcing
defined by the terms in the square brackets in eq~(\ref{f2n_onetime}).

We intend to study the small $\kappa$ limit of the problem. If
a given trajectory realization does not contain a `near hit' among
the trajectories, we expect that the $\kappa=0$ trajectories are in
some sense near the small $\kappa$ trajectories. The only case where
$\kappa=0$ differs radically from the small $\kappa$ limit is when
two (or more) trajectories become closer than
$O(\kappa^{1\over2-\xi})$. As we are interested only in separated
correlation
functions, we neglect the corrections to the $\kappa=0$ limit in what
follows.

All the existing studies of the Kraichnan model are based on analyzing
the Kraichnan linear PDEs. Most of these studies identify the anomalous
contributions as scale invariant functions of the coordinates which
are annihilated by the Kraichnan operator, thus they were termed zero
modes. Most researchers accept the picture that correlation functions
are obtained as a sum of zero modes with different anomalous scaling exponents
and a contribution which is directly related to the forcing which is
not anomalous. A notable exception to this approach is the Kraichnan closure,
based on an appealing but unproven conjecture on conditional
expectations of diffusive moments \cite{kr94}.

The common feature for these approaches is that the passive scalar is
treated as a field. Here, we propose an alternative viewpoint based on
eq.~(\ref{f2n}), in which
all the properties of low order moments are extracted by looking at a
small number of correlated stochastic trajectories.
The representation (\ref{f2n}) expresses $F_{2n}$ as
the expectation value of the forcing correlation function accumulated
over  the Lagrangian trajectories. Since the forcing drops off sharply
beyond  $L$ these
objects measure the correlation of the times for which the trajectory
positions are
composed of pairs of points which are within distance $L$.
In particular the two point moment $F_2$ is the average time during
which two
trajectories which end at given distance of each other stay within
distance $L$. Excluding the case $d=2,\,\xi=0$ the trajectories will
leave the vicinity of each other with probability 1, and the integrals
in~(\ref{f2n}) converge for each realization.

It should be noted that the representation described above can be used
in principle as a numerical scheme for computing directly the
correlation functions $F_{2n}$ using a Monte-Carlo method without the
need to generate the whole velocity field \cite{unpublished, massimo}.

In what follows we show how one can eliminate the forcing and the
outer scale from the problem. Anomalous scaling arises
from a relaxation process, in which the configuration tends towards an
asymptotic stationary distribution, while the overall scale
increases. The outer scale serves as a reference point
for this process, but is not intrinsic to the process itself.

\subsection{Shape evolution operator and its relation to multiscaling}

A component of the evolution of an initial configuration is a rescaling
of all the coordinates which all increase on the average like
$t^{1/\zeta_2}$; this rescaling is analogous to Richardson diffusion for the
present case \cite{legacy}.
The dynamic exponent $\zeta_2=2-\xi$  is also the
characteristic exponent of the second order structure function
\cite{kr68}. After
factoring  this overall expansion we are left with a normalized
`shape'. It is the evolution of this shape that determines the
anomalous scaling.

Consider an initial shape ${\Z}_{0}$  with an overall scale
$s_0$, that is, the configuration coordinates are
${\X}_{0}=s_0{\Z}_0$.
Note that the shape coordinates are
constrained. The overall scale is a homogeneous function of degree one
of the coordinates $s_{0}=S(\X_{0})$ but its exact functional form is to a
large degree arbitrary.

The  shape evolves in time. We fix a scale $s>s_0$ and examine
the  shape when the configuration  reaches the scale $s$ for the
first time (we know that this occurs in finite time with probability
one). Since the trajectories are random the new shape $\Z$ is taken
from a distribution $\gamma(\Z_{0},\Z,{s\over s_{0}})$. We used scale
invariance
to deduce that the distribution is a function of $s\over s_{0}$.

For a shape picked from an initial distribution of shapes $\rho_{s_0}(\Z_0)$
with scale $s_0$, the final shape distribution at scale $s$ is
\eq{inv}
\rho_{s}(\Z)=\int d \Z_{0} \rho_{0}(\Z_{0})\gamma(\Z_{0},\Z,{s\over s_{0}})
\en
For very large values of $s\over s_{0}$ we expect the final shape
distribution to approach an asymptotic distribution $\beta_{0}(\Z)$.
This distribution is invariant under the transformation (\ref{inv})
and is an eigenfunction of $\gamma$ with eigenvalue $1$.
In general we expect a spectrum of eigenfunctions $\beta_{n}$ for $\gamma$
with eigenvalues $\alpha_{n}({s\over s_0})$. By successive applications of
$\gamma$ it follows that
$\alpha_{n}({s\over s_0})=({s\over s_0})^{-\lambda_{n}}$.
Furthermore we assume that the eigenfunctions $\beta_{n}(\Z)$ form a
complete set. Thus, any initial distribution of shapes
$\rho_{s_0}(\Z)$ can be written as
\eq{eigenexpansion1}
\rho_{s_0}(\Z)=\sum_{n} A_{n} \beta_{n}(\Z)
\en
this distribution evolves into
\eq{eigenexpansion2}
\rho_{s}(\Z)=\sum_{m} A_{m}\left({s\over s_0}\right)^{-\lambda_{m}}
\beta_{m}(\Z)\ .
\en
Consider an average of a correlation function $F_{n}(\X)$ over $\rho_{s_0}$,
\eq{namely}
\left<F_{n}|s_{0}\right>=\int d\Z F_{n}(s \Z)\rho_{s_0}(\Z).
\en
We now use the representation~(\ref{f2n_onetime}) for
$F_n(s_0,\Z_0)$. We split the integration along the Lagrangian
trajectory at the first point where the overall scale
reaches the value $s$. It follows from the shape evolution
equation~(\ref{inv}) that we can write
\eq{split}
F_n(s_0,\Z_0)=I(s_0,s,\Z_0)+\int d\Z \gamma\left(\Z_0,\Z,{s\over s_0}\right)
F_n(s,\Z)\ .
\en
$I$ is the average of the contribution a specific forcing accumulated
along the path from $s_0$ to $s$. The forcing is that of
eq~(\ref{f2n_onetime}) so that for $s$ small enough its $L$ dependence
is only  through $F_{n-2}$.
 Averaging eq.~(\ref{split})
with $\rho_{s_0}(\Z_0)$ gives
\eq{diff}
\int d\Z [\rho_{s_{0}}(\Z)F_{n}(s_{0}\Z)-\rho_{s}(\Z)F_{n}(s\Z)]
=\left<I(s_0,s)\right>=
\sum_m A_{m}(\bar F_{n}^{(m)}(s_{0})-\left({s\over s_{0}}\right)^{\lambda_{m}}
\bar F_{n}^{(m)}(s))
\ .
\en
Where

\eqs
\bar F_{n}^{(m)}(s))=\int \beta_{m}(\Z)F_{n}(s\Z) d\Z
\en

and

\eqs
\left<I(s_0,s)\right>
=\int \rho_{s_{0}}(\Z_{0}) I(s_0,s,\Z_0) d\Z_{0}.
\en

The average $\left<I(s_0,s)\right>$ is dependent on $L$ only through
the behavior $F_{n-2}$ with $L$. Thus, this term cannot have
anomalous $L^{({n\over 2}(2-\xi)-\zeta_{n})}$  scaling.
Since the $A_{m}$'s are arbitrary, each term in the sum of
equation~(\ref{diff}) must be without anomalous
$L^{({n\over 2}(2-\xi)-\zeta_{n})}$  scaling as well.
The only way
for this to occur is if the anomalous $L$-dependent part of $\bar
F_{n}^{(m)}(s))$ is proportional to $f(L)s^{\lambda_{m}}$. Dimensional
considerations imply that $f(L)$ also has a power-law behavior that fixes
the dimension of $F_{n}$.

Thus, the projection of $F_{n}$ over $\beta_{m}$ selects a particular
anomalous component. We conclude that the exponents $\lambda_m$,
defined as the decay rates toward the invariant measure, are precisely
the anomalous exponents of $F_n$.

We can use the eigenfunctions $\beta_n$ to expand the operator $\gamma$
\eq{gamma}
\gamma(\Z_0,\Z,{s\over s_0})=\sum_m\left({s\over s_0}\right)^{-\lambda_{m}}
\beta_m(\Z) \mu_m(\Z_0).
\en
$\gamma$ is a non-Hermitian operator, and therefore we can expect only
that $\beta_m$ and $\mu_m$ are biorthogonal families of
functions. This means that if we expand $F_n$ in terms of the
functions $\mu_m$, a projection on $\beta_m$ will extract a single
term of the sum, which has a pure anomalous scaling exponent. This
leads us to identify the functions $\mu_m$ with the zero modes of the
Kraichnan operator, which were previously identified in
\cite{grisha,gk} as the anomalous scaling structures.

We will use this formalism in the next section in an explicit
perturbative calculation in large dimensions. This calculation
demonstrates all the concepts discussed in this section, and shows how
they can be used as a calculational method.

\subsection{Application: large $d$ perturbation theory}
The simplicity of the large dimension limit is due to the following reason:
random trajectories tend to separate faster the larger the dimension.
This is because there are more transversal directions in which to diffuse.
This property, which is well known for Brownian motion, is also
true for our case of correlated, Markovian paths. Therefore, in the limit of
infinite dimensions the motion becomes a deterministic growth of the
distance between each pair of points independently (note that the
infinite number of directions available makes such a motion possible).
This motion is simply the Richardson diffusion phenomenon, referred to
above, in which
relative distances $l$ increase like
\eq{rich} l\sim (l_0+t)^{1/(2-\xi)}. \en
This behaviour means that any initial $N$-point configuration will
evolve towards a regular $N-1$ dimensional simplex, where all the
relative distances are equal.

When $d$ is large but finite, trajectories fluctuate around the
deterministic infinite dimensional limit~(\ref{rich}), but the
fluctuations are small, of order $O(1/\sqrt d)$. This forms the basis
of perturbation theory in $1/d$.

Our procedure will be to examine the expectation value of a
non-dimensional symmetric combination of separations, which we denote
$\sigma$. The relaxation of $\left<\sigma\right>$
towards its asymptotic value is
described by the set of relaxation exponents $\lambda_m$,
see~(\ref{gamma})
\eq{sigma}
\left<\sigma|s\right>\equiv
\int d\Z\sigma(\Z)\gamma\left(\Z_0,\Z,{s\over s_0}\right)
=\sum_m \left({s\over s_0}\right)^{-\lambda_{m}} \bar\sigma_m
\mu_m(\Z_0), \en
where $Z_0$ is the initial shape, implying an initial value for
$\sigma(\Z_0)$. The $\bar\sigma_m$'s defined as
\eqs
\int \bar\sigma_m=d\Z\sigma(\Z)\beta_m(\Z),
\en
are numbers which only depend on the precise definition of $\sigma$. We see
from~(\ref{sigma}) that the functional dependence of
$\left<\sigma|s\right>$ on the initial shape $\Z_0$ is given by the
zero modes $\mu_m$. An asymptotic expansion in large $s\over s_0$ will thus
provide us with the leading zero modes, and their scaling exponents.

A term in the asymptotic expansion $\left<\sigma|s\right>$ is expected
to be, in the large $d$ limit, of the form
\eq{pllog}
\left({s\over s_0}\right)^{-\lambda^{(0)}-{1\over d}\lambda^{(1)}+\cdots}[
\mu^{(0)}(\Z_0)+{1\over d}\mu^{(1)}(\Z_0)+\cdots].
\en
Expanding the exponent gives
\eq{llog}
\left({s\over s_0}\right)^{-\lambda^{(0)}}[\mu^{(0)}(\Z_0)+
{1\over d}(-\lambda^{(1)}\log({s\over s_0})+\mu^{(1)}(\Z_0))+\cdots].
\en
Hence, the correction to the scaling exponent can be read off by
looking at the logarithmic term in $1\over d$. Although nominally our
expansion is in $1\over \sqrt{d}$  it turns out that the first term
vanishes (see below).

In principle, the statement that a logarithmic contribution is the
first term in an expansion of a power needs to be justified by showing
that all higher order terms of order ${1\over d^n}\log({s\over s_0})^{n}$ are
consistent with the first term. In our case this is not
necessary. Rather, we consider a perturbative expansion for the
equivalent Fokker-Planck equation. In the Fokker-Planck equation the
eigenvalues
play the role of the exponents $\lambda_{m}$. Since perturbation
theory finds corrections in the eigenvalues, we conclude that the logarithms
must sum to a pure power as expected.

The perturbative analysis is going to carried out in terms of time as
an independent variable rather than $s$. We thus need at some stage to
transform to the $s$ variable in which our theory is formulated.
Since the dynamics are nearly deterministic, for a given $s$, $t$ is
very narrowly distributed around its Richardson diffusion value $t\sim
s^{2-\xi}$. We show that to the order that was retained in the
perturbation calculations it is sufficient to replace $t$ by its
Richardson diffusion value.

\subsubsection{Representation in terms of separations, and
perturbation theory}
The basic variables we use for analysis in high dimensions are the
$n_s\equiv N(N-1)/2$ inter-point square separations
$\tilde q_{nm}\equiv(\r_n-\r_m)^2$. Their time evolution is
obtained from (\ref{basic3}) by the rules of the Ito calculus
\eq{basics} {d\tilde q_{nm}(t)\over dt}=\left<(\U_n-\U_m)^2\right>+
2(\r_n-\r_m)\cdot(\U_n-\U_m)=2{\tilde q_{nm}(t)^{\xi/2}\over\ell^\xi}
(d-{\xi\over d-1+\xi}) +\tilde\eta_{nm}(t), \en
where the second equality serves to define the noise term
$\tilde\eta_{nm}$, a zero mean, $\delta$-correlated in time Gaussian
process. The equations are again to be interpreted as Ito SDEs.
The $n_s$ separations are subject to triangle inequalities
which are preserved by the dynamics (\ref{basics}).

The deterministic part in eqs.~(\ref{basics}) gives simple Richardson
diffusion with the exponent $2/(2-\xi)\equiv2/\zeta_2$, whereas all
the non-trivial behavior is contained in the $\tilde\eta$ terms which
couple between the different separations. When $d$ is large the
noise term in (\ref{basics}) becomes small with respect to the
deterministic term, and this phenomenon serves as the starting point for
perturbation theory. In order to demonstrate this it is convenient to
factor out the super-diffusive behavior from $\tilde q_{nm}$, defining
\eqs \tilde q_{nm}(t)=\left({2-\xi\over\ell^\xi}d(t+\tau_{nm})\right)
^{2/(2-\xi)}q_{nm}(t), \en
where $\tau_{nm}$ is a constant determined by initial
conditions. Substituting in eq.~(\ref{basics}) yields (here and in the
following we neglect the higher order term $\xi/(d-1+\xi)$ coming from
incompressibility)
\eq{fluct} \dot q_{nm}={2\over2-\xi}{1\over t+\tau_{nm}}
\left(q_{nm}^{\xi/2}-q_{nm}\right)+{1\over\sqrt d}\eta_{nm}(t),\en
where
\eqs\eta_{nm}=\left({2-\xi\over\ell^\xi}d(t+\tau_{nm})\right)^{-2/(2-\xi)}
\tilde\eta_{nm}\ .\en
The covariance of $\eta_{nm}$ is (using the abbreviation
$\bar q_{nm}=(t+\tau_{nm})^{2/(2-\xi)} q_{nm}$)
\eq{etacov3} \begin{array}{l}\displaystyle
\left<\eta_{nm}(t)\eta_{kl}(t')\right>={2\over\zeta_2}
(t+\tau_{nm})^{-2/\zeta_2}(t+\tau_{nl})^{-2/\zeta_2}\\\displaystyle\qquad
\qquad\times
(\bar q_{nl}-\bar q_{ml}-\bar q_{nk}+\bar q_{mk})
(\bar q_{nl}^{1-\zeta_2/2}-\bar q_{ml}^{1-\zeta_2/2}-
\bar q_{nk}^{1-\zeta_2/2}+\bar q_{mk}^{1-\zeta_2/2}) \delta(t-t')\ .
\end{array}\en
This covariance is of order 1 in $d$,
so that the stochastic term in eq.~(\ref{fluct}) is indeed small for
$d\gg1$. Thus, the dynamics of eq.~(\ref{fluct}) are dominated by the
attractive fixed point $q_{nm}=1$ of the deterministic term, and the
random term causes small fluctuations around it.

Perturbation theory is performed in a straightforward manner by
expanding
\eq{spert}
 q_{nm}=1+{1\over\sqrt d}q_{nm}^{(1)}+{1\over d}q_{nm}^{(2)}+\ldots\ ,\en
and
\eqs \eta_{nm}={1\over\sqrt d}\eta_{nm}^{(1)}+
{1\over d}\sum_{kl}\eta_{nm,kl}^{(2)}q_{kl}^{(1)}+\ldots\ .\en
The noise terms are white Gaussian zero-mean processes. The
correlation matrix of $\eta_{nm}^{(1)}$, is obtained
from~(\ref{etacov3}) by substituting $s_{nm}\rightarrow1-\delta_{nm}$.
The correlations involving $\eta_{nm}^{(2)}$ are more complicated, but
are not needed to the order that we keep in the calculations.

The terms in the perturbation series~(\ref{spert}) obey linear
inhomogeneous equations:
\eqs \dot q_{nm}^{(1)}=-{1\over t+\tau_{nm}}q_{nm}^{(1)}+\eta_{nm}^{(1)}\ ,\en
and
\eqs \dot q_{nm}^{(2)}=-{1\over t+\tau_{nm}}
(q_{nm}^{(2)}+{\xi\over4}(q_{nm}^{(1)})^2)+
\sum_{kl}\eta_{nm,kl}^{(2)}q_{kl}^{(1)}\ ,\en
with respective solutions

\eq{s1} q_{nm}^{(1)}(t)={1\over t+\tau_{nm}}\int_0^tdt'(t'+\tau_{nm})
\eta_{nm}^{(1)}(t')\ ,\en
and
\eq{s2} q_{nm}^{(2)}(t)={1\over t+\tau_{nm}}\int_0^tdt'\left[
-{\xi\over4}
q_{nm}^{(1)}(t)^2
+(t'+\tau_{nm})\sum_{kl}\eta_{nm,kl}^{(2)}(t')q_{kl}^{(1)}(t')\right]\ .\en

\subsubsection{Results for three-and four-point functions}
In this section we are going to use the results of perturbation theory
derived above to demonstrate multiscaling in the three- and four-point
correlation functions. For this purpose we make the following
(arbitrary) choice
\eqs \sigma\equiv n_s{\sum_{nm} \tilde q_{nm}^2\over
(\sum_{nm} \tilde q_{nm})^2}.\en
The definition of the overall scale is chosen to be
\eqs s=\sqrt{{1\over N}\sum_{n<m}\tilde q_{nm}}\ . \en
It is shown in the following subsection that replacing $t$ by
$s^{\zeta_2}$ is correct to the order which we keep in the following
calculation.

Substituting the perturbation series~(\ref{spert}) to order $1/d$
gives the expansion $\sigma=\sigma^{(0)}+{1\over \sqrt{d}}\sigma^{(1)}
+{1\over d}\sigma^{(2)}+\cdots$, where we have defined
\eq{sigma0}\sigma^{(0)}=n_s{\sum_{nm} \rho_{nm}^{4/\zeta_2}\over\left(\sum_{nm}
\rho_{nm}^{2/\zeta_2}\right)^2}\ ,
\en

\eq{sigma1}\sigma^{(1)}=\sigma^{(0)}
\left[2{\sum_{nm} \rho_{nm}^{4/\zeta_2}q^{(1)}_{nm}\over\sum_{nm}
\rho_{nm}^{4/\zeta_2}}-2{\sum_{nm}\rho_{nm}^{2/\zeta_2}q^{(1)}_{nm}
\over\sum_{nm}
\rho_{nm}^{2/\zeta_2}}\right] \en

\eq{sigma2}\begin{array}{l}\displaystyle\sigma^{(2)}=\sigma^{(0)}\Bigg[
{\sum_{nm} \rho_{nm}^{4/\zeta_2}(q^{(1)}_{nm})^2\over\sum_{nm}
\rho_{nm}^{4/\zeta_2}}+2\left(
{\sum_{nm} \rho_{nm}^{4/\zeta_2}q^{(2)}_{nm}\over\sum_{nm}
\rho_{nm}^{4/\zeta_2}}-{\sum_{nm}
\rho_{nm}^{2/\zeta_2}q_{nm}^{(2)}\over\sum_{nm}
\rho_{nm}^{2/\zeta_2}}\right)\\\qquad\qquad\displaystyle+
3\left({\sum_{nm} \rho_{nm}^{2/\zeta_2}q^{(1)}_{nm}\over\sum_{nm}
\rho_{nm}^{2/\zeta_2}}\right)^2-
4{\sum_{nm} \rho_{nm}^{4/\zeta_2}q^{(1)}_{nm}\over\sum_{nm}
\rho_{nm}^{4/\zeta_2}}{\sum_{nm}
\rho_{nm}^{2/\zeta_2}q^{(1)}_{nm}\over\sum_{nm}
\rho_{nm}^{2/\zeta_2}}\Bigg]\ .\end{array} \en
(We used the abbreviation $\rho_{nm}=t+\tau_{nm}$.)

When $t\gg\tau_{nm}$, $\left<\sigma\right>$ is dominated by
contributions from the leading zero modes, and we therefore check its
asymptotics in this limit. In the limiting case
$d=\infty$ we have, for $t\rightarrow\infty$
\eq{sg0}\begin{array}{r}\displaystyle \left<\sigma\right>=\sigma^{(0)}\sim
n_s{\sum_{nm}1+(4/\zeta_2)(\tau_{nm}/t)+(2/\zeta_2)(4/\zeta_2-1)(\tau_{nm}/t)^2
\over(\sum_{nm}1+(2/\zeta_2)(\tau_{nm}/t)
+(1/\zeta_2)(2/\zeta_2-1)(\tau_{nm}/t)^2)^2}
\qquad\\\displaystyle
\sim 1+{4\over n_s\zeta_2^2}\sum_{nm}\left({\tau_{nm}\over t}\right)^2-
{4\over(n_s\zeta_2)^2}\left(\sum_{nm}{\tau_{nm}\over t}\right)^2\ .
\end{array}\en
In the special case $N=3$, $n_s=3$ eq.~(\ref{sg0}) can be written as
\eq{sigma0-f} \sigma^{(0)}\sim 1+{8\over9\zeta_2^2}{Z_3\over t^2}
\en
where $Z_3$ is the leading three-point zero mode,
\eq{z3} Z_3=\sum_{n\ne m,l\atop m<l}(\tau_{nm}-\tau_{nl})^2 \en
$\sigma^{(0)}$ is thus a
linear combination of zero modes (up to order $t^{-2}$) as expected.

The $O(1/\sqrt d)$ contribution $\left<\sigma^{(1)}\right>$ vanishes since
$\left<\eta^{(1)}\right>=0$, so that we proceed to examine the
$O(1/d)$ term. Up to order $t^{-2}$, $\left<\sigma\right>$ contains,
in addition to the constant term, a term which is proportional to
$\log t/t^2$. This term reflects the dependence of the scaling
exponent on $d$, and we will therefore display only this term
explicitly.

The first step consists of calculating moments of the $s$ functions:
\eqs \begin{array}{r}\displaystyle
\left<q_{nm}^{(1)}(t)^2\right>={1\over(t+\tau_{nm})^2}
\int_0^t dt' dt'' (t'+\tau_{nm})(t''+\tau_{nm}) \left<\eta_{nm}^{(1)}(t)
\eta_{nm}^{(1)}(t')\right>\qquad\\\displaystyle
={4\over\zeta_2}\left(1-{\tau_{nm}^2\over(t+\tau_{nm})^2}
\right)\ .\end{array}\en

\eq{q2}\left<q_{nm}^{(2)}(t)\right>={1\over(t+\tau_{nm})}(-{\xi\over4})
\int_0^tdt \left<q_{nm}^{(1)}(t')^2\right>=-{2-\zeta_2\over\zeta_2}
\int_0^t dt' \left(1-{\tau_{nm}^2\over(t'+\tau_{nm})^2}\right)\ .\en
(The term proportional to $\eta^{(2)}$ drops in averaging due to the Ito
convention). Thus, terms involving only one separation do not
contribute logarithmic terms to this order.

The large $t$ asymptotics of the cross correlation of two separations with
one common vertex is
\eqs\begin{array}{l}\displaystyle
 \left<q_{nm}^{(1)}(t)q_{nl}^{(1)}(t)\right>
\sim
 c_{0nml}+{c_{1nml}\over t}+\\\qquad\displaystyle
{2\over\zeta_2}{\log t\over t^2}\left({2\over\zeta_2}-1\right)
\bigg[{1\over\zeta_2}(\tau_{nm}^2+\tau_{nl}^2)+\left({2\over\zeta_2}-1\right)
(\tau_{nm}^2+\tau_{nl}^2-\tau_{ml}^2)+\left({2\over\zeta_2}-1\right)
\tau_{nm}\tau_{nl}
\\\qquad\displaystyle-\left({4\over\zeta_2}-1\right)(\tau_{nm}+\tau_{nl})
(\tau_{nm}+\tau_{nl}-\tau_{ml})+{2\over\zeta_2}
(\tau_{nm}+\tau_{nl}-\tau_{ml})^2\bigg]\ .\end{array}\en
Summing over permutations gives
\eq{sigma2-p} \begin{array}{l}\displaystyle
\sum_{n\ne m,l\atop m<l}\left<q_{nm}^{(1)}(t)q_{nl}^{(1)}(t)\right> \sim
\tilde c_{0}+{\tilde c_{1}\over t}+{2\over\zeta_2}{\log t\over t^2}
\left({2\over\zeta_2}-1\right) \left({2\over\zeta_2}+1\right)Z_3\ .
\end{array}\en

We conclude that the only terms in $\left<\sigma^{(2)}\right>$ which
generate logarithms are
the ones involving cross correlations of $q$ variables, {\it i.e.},
the last two in eq.~(\ref{sigma2}). Of these terms we need only keep the
leading order in $1/t$, giving
\eq{sigma2-f}\left<\sigma^{(2)}\right>=c_{0}+{c_{1}\over t}-
{4\over9\zeta_2}
\left({2\over\zeta_2}-1\right)
\left({2\over\zeta_2}+1\right)Z_3{\log t\over t^2}\ .\en

Combining the contributions from~(\ref{sigma0-f}) and~(\ref{sigma2-f})
yields
\eq{3point} \left<\sigma\right>\sim1+{8\over9\zeta_2^2}{Z_3\over t^2}
(1-{\xi\over2}\left({2\over2-\xi}+1\right){\log t\over d})\ .\en
which implies, after substituting $t\rightarrow s^{2-\xi}$
\eqs \zeta_3=2(2-\xi)+
{\xi\over2d}\left(4-\xi
\right) +O\left({1\over d^{3/2}}\right)\ .\en

A similar analysis is needed for the case $N=4$, $n_s=6$. The infinite
dimensional limit becomes in this case
\eq{s0-4p} \sigma^{(0)}\sim
1+{1\over9\zeta_2^2}{Z_3+Z_4\over t^2}\ ,\en
where the leading order 4-point zero mode is
\eqs Z_4=\sum_{n<m,\,k<l\atop n<k} (\tau_{nm}-\tau_{kl})^2\ .\en
We also need cross correlations of separation variables without a
common vertex, whose large $t$ asymptotics are
\eq{s21-4p}\begin{array}{r}\displaystyle
\left<q_{nm}^{(1)}q_{kl}^{(1)}\right>
\sim \left({2\over\zeta_2}\right)^2\left({2\over
\zeta_2}-1\right){\log t\over t^2}\left(\tau_{nl}-\tau_{ml}-\tau_{nk}+\tau_{mk}
\right)^2\ ,\end{array}\en
and summing over permutations gives
\eq{s22-4p} \sum_{n<m,\,k<l\atop n<k} q_{nm}^{(1)}q_{kl}^{(1)}\sim
\left({2\over\zeta_2}
\right)^2\left({2\over\zeta_2}-1\right){\log t\over t^2} (-2Z_4+Z_3)\ .\en
As in the 3-point case we collect the contributions
from~(\ref{s0-4p}), (\ref{s21-4p}), and~(\ref{s22-4p}) giving
\eq{4point} \left<\sigma\right>\sim1+{1\over9\zeta_2^2}{1\over t^2}
\left(Z_3+Z_4+{\log t\over d}\xi\left[-\left({3\over2\zeta_2}+{1\over2}\right)
Z_3+{4\over\zeta_2}Z_4\right]\right)\ ,\en
and the dependence on $Z_4$ implies that
\eqs \zeta_4=2(2-\xi)-{4\xi\over d}+O({1\over d^{3/2}}),\en
in agreement with~\cite{grisha}.
Eq.~(\ref{4point}) yields further information by checking the contributions
proportional to $Z_3$; the relative strength of the logarithmic terms
is different from that obtained in the three point analysis,
see~(\ref{3point}). The reason is that $Z_4$ generates contributions
proportional to $Z_3$ at first order, so that it is no longer an
approximate zero mode when $d$ is finite, and the degeneracy between
the 3-point and 4-point zero modes breaks down. It is not difficult to
build the 4-point zero mode $\bar Z_4$ which continues correctly from
infinite to finite dimensions as a linear combination of $Z_3$ and $Z_4$
\eqs \bar Z_4=Z_4+{4\over12-\xi}Z_3 .\en

\subsubsection{The \protect{$t\rightarrow s$} transformation}
The previous analysis relied on calculations as a function of time,
and the transformation to representation as a function of total scale
was carried out by the simple substitution $t\rightarrow s^{\zeta_2}$.
The purpose of this subsection is to show that corrections due to
non-trivial dependence of $s$ on $t$ do not contribute terms of the
order
which was kept above, {\it i.e.} $O(1/t^2)$ and $O(\log t/(d t^2))$.

It should be emphasized that even when $d$ is very large, $s$ is not a
monotonic function of $t$, and this is evident even in first order
perturbation theory. However, the size of the time interval in which one
is likely to find the same of value of $s$ becomes very small. In order
to estimate the width of this interval, suppose
$s(\bar t)=\bar s$. We ask for which values of $\tau$ it is likely that
$s(\bar t+\tau)=\bar s$ as well. We know that for small $\tau$
\eqs s(\bar t+\tau)\sim \bar s+ C \tau+ {B\over\sqrt d}\sqrt\tau
\omega, \en
where $\omega$ is a random variable with $O(1)$ variance, and $B$ and
$C$ are some order 1 numbers. The deterministic part is monotonic and
may be compensated by the fluctuating part to give a solution only if
$\tau=O(1/d)$. This is our estimate for the width of the multiple
solution region. Replacing the earliest solution for $t(s)$ by some
other arbitrary solution thus induces an $O(1/d)$ error in $t$, which
is too small to affect the results to the order that we keep.

In addition, up to $O(1/d)$ we may assume that $s(t)$ is a one-to-one
mapping, and may invert the relation to obtain $t(s)$.
$s$ may be written (up to a constant multiplier) as
\eqs s(t)
\sim\sqrt{{1\over N}\sum_{nm}
(t+\tau_{nm})^{2/\zeta_2}(1+{1\over\sqrt d}q_{nm}^{(1)}(t)+\cdots)}\ .\en
Inverting to express $t$ as a function of $s$ gives
\eq{tofs} t=t_0(s)\left(1-{1\over t_0(s)}{2\over\zeta_2}\sum_{nm}\tau_{nm}-
{1\over\sqrt d}\sum_{nm}q_{nm}^{(1)}(t_0(s))\right)
+\mbox{higher order terms},\en
where $t_0=s^{\zeta_2}$. We see that to $O(1/\sqrt d)$, $t$ may indeed
be considered as a single valued function of $s$ which is however
randomly shifted with respect to the zero order estimate. We are now
going to show that this correction does not contribute to the
perturbative results either.

We substitute $t(s)$ in the asymptotic expression for $\sigma^{(0)}$ (see
eq.~\ref{sg0}, keeping the dependence on $\tau_{nm}$ implicit),
\eqs \sigma^{(0)}\sim1+{F(\tau_{nm})\over t^2}\sim1+{F(\tau_{nm})\over t_0^2}
\left(1+{2\over t_0}{2\over\zeta_2}\sum_{nm}\tau_{nm}+
{2\over\sqrt d}\sum_{nm}q_{nm}^{(1)}(t_0)+{3\over d}
(\sum_{nm}q_{nm}^{(1)}(t_0))^2\right)\ .\en
It follows after averaging that corrections to $t_0$ will contribute
only terms of $O(1/t_0^3)$ and $O(\log t_0/(d t_0^4))$.

$\left<\sigma^{(1)}\right>$,
which vanishes when one takes $t=t_0$ as in
the previous subsection, is non-zero when corrections to $t_0$ are
taken into account. The leading non-zero terms are proportional to
\eqs {1\over\sqrt d}\sum_{nm} {\tau_{nm}q_{nm}^{(1)}\over t_0}
\sum_{kl}q_{kl}^{(1)}(t_0)\ ,\en
whose expectation is of order $O(1/(\sqrt d t_0))$ and
$O(d^{-3/2}\log t_0/t_0^3)$.
Finally, it is clear that corrections to $t_0$
in $\left<\sigma^{(2)}\right>$ cannot change the the leading
logarithmic behavior. We conclude that to the order to which the
above calculations we carried out it is safe to take $t=t_0=s^{\zeta_2}$.

\subsection{Conclusions and further applications}
In this paper we have presented a path oriented approach to the study
of passive scalar advection. The main difference between this approach
and other studies is that we do not need to consider the whole flow
(and scalar) field, rather, an $N$-point correlation function is
described via the evolution of $N$ Lagrangian trajectories. This allows
us to study ordinary differential equations (albeit stochastic ones)
instead of partial differential operators. The anomalous scaling is
due to relaxation towards an invariant distribution of the
instantaneous shapes. Since the relaxation rates are associated with
excited states of an evolution operator these scaling exponents are
not related to the normal scaling and dimensional reasoning cannot be
applied.

The Lagrangian trajectories become nearly deterministic in very large
dimensions (up to random solid body rotations). This property explains
why the dynamics simplifies in this limit, and also serves as a
starting point for an expansion. Using perturbation theory all the
ingredients of the new anomalous scaling picture were demonstrated in
an explicit and concrete manner, and perturbative corrections to the
anomalous exponents were calculated.

In addition to the perturbative application it is possible to use the
same concepts for numerical Monte-Carlo simulations. The main
numerical task is to generate the random trajectories, which can be
done using standard methods. However, in preliminary studies severe
problems of convergence prevented us from obtaining precise results,
and therefore the presentation of the numerical application is
postponed \cite{unpublished,massimo}.

It would also be interesting to see how the other integrable limits of
this model ($\xi\rightarrow0$ and $\xi\rightarrow2$) appear
using the present ideas.

We would like to thank Uriel Frisch, Zeev Olami, Itamar Procaccia
and
Massimo Vergassola for discussions. O.G. thanks Krzystof Gawedzki and
Itamar Procaccia, the organizers of the Turbulence Workshop in I.H.E.S
in April 1997, where many of the ideas in this paper were developed.
Most of this work was done while R.Z. was a CNRS post-doctoral fellow
in the Laboratoire de Physique Statistique at the Ecole Normale
Superieure (Paris). The Laboratoire de Physique Statistique is
Laboratoire associ\'e aux Universit\'es Pierre et Marie Curie,
Denis Diderot et au CNRS.



\begin{thebibliography}{99}

\bibitem{legacy} U. Frisch, {\it Turbulence}, Cambridge University
Press (1995).
\bibitem{kr68} R. H. Kraichnan, Phys. Fluids {\bf11} 945 (1968)
\bibitem{kr94} R. H. Kraichnan, Phys. Rev. Lett., {\bf 72} 1016 (1994)
\bibitem{gk} K. Gaw\c{e}dzki and A. Kupiainen, Phys. Rev. Lett., {\bf
75},  3834, (1995).
\bibitem{grisha} M.Chertkov, G. Falkovich, I. Kolokolov and V. Lebedev,
Phys. Rev. E {\bf 52} 4924, (1995).
\bibitem{3pt} O. Gat, V.S. L'vov, and
I. Procaccia, Phys. Rev. {\bf E56}, 406 (1997).
\bibitem{slow} A similar point of view is presented in D. Bernard,
K. Gawedzki, A. Kupiainen, e-print cond-mat/9706035 (1997).
\bibitem{drummond} See for example, I. T. Drummond, J. Fluid Mech.
{\bf123}, 59 (1982), M. Chertkov, Phys Rev. E {\bf55}, 2722, (1997),
P. M. Ginanneschi, e-print no. chao-dyn/9704017 (1997).
\bibitem{dz} B. Derrida, R. Zeitak, Phys. Rev. E. {\bf54} pp 2513 (1996).
\bibitem{kyc} R.~H. Kraichnan, V. Yakhot and S. Chen,
Phys. Rev. Lett., {\bf 75}, 240 (1996).
\bibitem{unpublished} O. Gat, I. Procaccia, and R. Zeitak, unpublished
(1997).
\bibitem{massimo} The idea of using paths for passive scalar
Monte-Carlo simulation was developed independently by U. Frisch and
M. Vergassola.
%
\end{thebibliography}
\end{document}